 \documentstyle[11pt]{article}
\def\ll{\label}
\def\re{\ref}
\def\c{\cite}

\def\r1{(\ref{$1})}

\def\ba{\begin{array}{c}}

\def\ea{\end{array}}

\def\ni{\noindent}
\def\si{\sigma}

\def\De{\Delta}

\def\ov{\over}
\def\ha{{1\over 2}}

\def\l{\left}
\def\l({\left(}
\def\r){\right)}
\def\r{\right}
\def\rw{\rightarrow}

\def\la{\lambda}
\def\al{\alpha}

\def\be{\begin{equation}}
\def\bc{\begin{center}}
\def\ec{\end{center}}
\def\bit{\begin{itemize}}
\def\eit{\end{itemize}}
\def\ee{\end{equation}}
\def\ed{\end{document}}
\def\bea{\begin{eqnarray}}
\def\eea{\end{eqnarray}}
\def\efr{\end{flushright}}

\begin{document}
\title{Generation of  new classes of   integrable quantum and
statistical models
}

\author{
Anjan Kundu \footnote {email: anjan@tnp.saha.ernet.in} \\  
  Saha Institute of Nuclear Physics,  
 Theory Group \\
 1/AF Bidhan Nagar, Calcutta 700 064, India.
 }
\maketitle
\vskip 1 cm

\begin{abstract} 

A scheme based on a 
unifying  q-deformed algebra and associated with a 
 generalized Lax operator is proposed for generating
   integrable quantum and statistical models.  As important applications we
 derive known as well as novel quantum models and
   obtain new series of  vertex models related to q-spin, q-boson and
    their hybrid combinations. Generic q,  q roots of unity and $q \to 1$
 yield different classes of integrable models.
     Exact solutions through algebraic Bethe ansatz is formulated for all
models in a unified way.

-----
\medskip

 PACS numbers 02.30.Ik,
  02.20.Uw,
05.50+q,
03.65.Fd

\medskip

\end{abstract}

\smallskip

  
\section{Introduction}
\setcounter{equation}{0}
Integrable quantum systems in $1+1$-dimensions and statistical models in
$2$-dimensions belong to an exclusive and important class of models giving
exact results. However there seems to be no established
  scheme for generating all such models along with their Lax operators and
 $R$-matrices in a unified way. Our aim therefore is an
effort in that direction for constructing a  wide range of known as well as
new classes of models putting particular emphasis on statistical models.

The integrable systems may be defined by the  property that they
possess N-number of conserved quantities, which are independent and commute
among themselves: $[c_n,c_m]=0$, where N is to the degree of freedom of
the system. Such integrability also leads to the exact solvability of the
model, since taking conserved quantities as action variables
 one can adopt the
action-angle variable   
 description.
For managing conveniently   such a rich structure 
 we  introduce  first a generating function 
$\tau(\lambda)$ depending on some extra  parameter $\la$, known as 
the  spectral parameter,
 such that one can recover the infinite number of conserved quantities
including the Hamiltonian as
the expansion coefficients of $\ln \tau(\lambda) = \sum_jc_j\lambda^j$.
The crucial integrability condition may then be defined in a compact form as
$[\tau(\lambda),\tau(\mu)]=0$, from which  the commutativity of $c_j$'s
follows immediately by comparing the coefficients of different powers of
$\la , \mu$. The partition function $Z$
 of the related statistical 
 model on the other hand can be  constructed from  $\tau(\la)$ as $Z=
tr(\tau(\la)^M)$. 

To be convinced of  the amazingly wide   variety and range of 
the  integrable  models,
 let us  name a few of them \c{kulskly},
  which allow both classical and quantum variants.
 They are field and lattice models of relativistic
or  anisotropic  type such as  sine-Gordon   field 
and  exact  lattice \c{lsg} model  and similarly
Liouville model,
 quantum derivative nonlinear Schr\"odinger    
 (DNLS) equation \c{qdnls} together with their exact  lattice versions,
  quantum spin model like $XXZ$  chain \c{6v},  
relativistic quantum  Toda chain \c{qrtc} etc.  Models can also be
of nonrelativistic or isotropic type such as
 NLS model \c{kulskly} and its lattice variant,
 $XXX$ spin chain \c{6v}, 
 Toda chain \c{kulskly} etc.  Among the statistical integrable models
the most simple and well known is the {\it 6-vertex model} \c{baxter}. This
model is defined on a square lattice with a random direction on
 each bond ( left or  right on horizontal, up or down on  vertical)
(FIG.1a)) 
constrained by the {\it ice rule},
 that the number of incoming and outgoing arrows at each vertex are the
same. This leaves only $6$ possible configurations giving that many
 Boltzmann weights (BW) $\omega_{i,j;k,l}$
 at each vertex point, coining the name of
the model. These $6$  weights may be given by the nontrivial matrix
elements of the $R$-matrix (\re{R-mat}) and 
the  partition function  is expressed  as 
 $Z= \sum_{config} \prod_{a,b,j,k} \omega_{a,j;b,k}(\la).$

\section{Lax operator approach}
For describing the intricate  
  integrable structures mentioned above  
one can not  however start just
from the Hamiltonian of the model as conventional  in
physics, since now the Hamiltonian is merely one among many commuting
conserved charges. We on the other hand have to  allow  certain
abstractions and start with an unusual type of matrix called the Lax operator
$L_j(\la)$ defined at each site $j$ of a 1-dimensional discretized lattice.
The matrix elements of the Lax operator, unlike in an  usual matrix,
 are themselves operators acting on
some Hilbert space. 
For ensuring the integrability of a system the related 
$L_j(\la)$ generally   
should satisfy certain commutation relations given by the 
ultralocality condition
 $L_j(\la)\otimes  
L_k(\mu)= (I \otimes L_k(\mu) ) \otimes ( L_j(\la)\otimes I) 
$
 at different lattice points  and the Yang-Baxter equation  
\be
R(\la-\mu) L_j(\la)\otimes  
L_j(\mu)= (I \otimes L_j(\mu) ) \otimes ( L_j(\la)\otimes I) R(\la-\mu).
\ll{ybe}\ee
 at the same points $j=1,2, \ldots, N.$  The $R(\la-\mu) $-matrix
has spectral parameter dependent
 c-number elements, which we take as a $4\times 4$ matrix 
  with
\begin {equation} 
R^{11}_{11} = R^{22}_{22}= \sin \al (\la +1) ,
\  R^{12}_{12} = R^{21}_{21}= \sin \al (\la ), \ R^{12}_{21} =
 R^{21}_{12}= \sin \al . 
        \ll{R-mat}\end {equation}
Since our intension is to establish the integrability which is a global
property, we have to switch from this local picture at each site $j$ to some
global one by defining a matrix $
 \ T(\la)=\prod_{j=1}^N L_J(\la) $. Multiplying therefore the YBE
(\re{ybe}) for $j=1,2, \ldots, N$ and thanks  to the ultralocality condition 
  one arrives at the global YBE:
$~~ R_{12}(\lambda - \mu)~ { T} (\lambda)~\otimes { T}(\mu )= (I\otimes 
{ T}(\mu )~ {
~  T\otimes I }(\lambda)~~ R_{12}(\lambda - \mu)$
having exactly the same structural form. Invariance of the same algebraic
form also for the tensor product of the algebras as revealed here exhibits
an
 underlying deep Hopf algebra structure. Defining further 
$~\tau(\la)=tr T(\la)$, taking $trace$ from both sides of the global YBE
and canceling  the $R$-matrices due to the cyclic rotation of
matrices under the  trace  we arrive finally at the 
trace identity through $~\tau(\la) $ defining 
the quantum integrability of the system.

Therefore we see that the integrable systems may be defined by their Lax
operators satisfying the ultralocality condition and the YBE together with
the associated $R$-matrix. Note that  all the varied integrable 
models listed above  have their representative $L$ operators as $2\times 2$
matrices, but with diverse 
 forms expressed through different basic operators like 
 spins  $\vec \sigma$, bosons 
 $\psi,\psi^\dag$ or
real canonical operators  $u,p$ etc. However 
surprisingly, the $R$-matrix  associated with all of them  appears to be 
  the
 same and given by the 
 known trigonometric solution (\re{R-mat}) as for the 6-vertex model
 (or its $q = e^{i\al}\to 1$ ,
i.e rational limit).
Therefore a natural question arises asking  whether 
  there can be a {\it unified} model associated with 
(\re{R-mat}) such that all above models can be derived from it in
a systematic way.

\section {Unified model}
We show  that it is indeed  possible to find such an integrable
 unified model
and justify  consequently the
 sharing of the same
$R$-matrix by all  other integrable models, which will be 
obtained   through various reductions and realizations of this
single model. 
 Such a general model must be a
quantum, lattice model, which is relativistic or anisotropic and extendible
to be inhomogeneous with inbuilt quantum parameter $\hbar$, lattice constant
$\Delta$,   
deformation
parameter $q$ and the
inhomogeneity parameters $\{c\}$), such that it can generate lattice/ field
models, quantum/ classical models, relativistic (anisotropic)/
nonrelativistic (isotropic) variants and similarly inhomogeneous/homogeneous
models
 at the corresponding  limits of
these parameters.

Since the integrable models can be  represented
  by their   Lax
operators, we propose the  defining form for our unified model as
\be
L_t^{(anc)}{(\xi)} = \left( \begin{array}{c}
  \xi{c_1^+} q^{ S^3}+ \xi^{-1}{c_1^-}  q^{- S^3}\qquad \ \ 
 \epsilon  S^-   \\
    \quad  
  \epsilon S^+    \qquad \ \  \xi{c_2^+}q^{- S^3}+ 
\xi^{-1}{c_2^-}q^{S^3}
          \end{array}   \right), \quad
         q=
e^{i \alpha } , \ \xi=e^{i \alpha \la}, \epsilon= 2\sin \al . \ll{nlslq2} \ee
with
the basic operators satisfying the quadratic algebra 
\be
 [S^3,S^{\pm}] = \pm S^{\pm} , \ \ \ [ S^ {+}, S^{-} ] =
 \left ( M^+\sin (2 \al S^3) + {M^- } \cos
( 2 \al S^3  ) \right){1 \over 2\sin \al}, \quad  [M^\pm, \cdot]=0,
\ll{nlslq2a}\ee
where  $ M^\pm$ are  the central elements expressed as  
  $ M^\pm=\pm   \sqrt {\pm 1} ( c^+_1c^-_2 \pm
c^-_1c^+_2 ) $ through the commuting set of elements $c_a^\pm$.
The algebra (\re{nlslq2a}) is a novel q-deformed 
 algebra and includes 
known q-spin as well as q-boson algebras as   particular reductions.
 The quantum integrability of this model and hence all other models
derived below from it is guaranteed, since (\re{nlslq2}) with (\re{nlslq2a})
associated with 
$R$-matrix (\re{R-mat}) satisfy YBE (\re{ybe}).

It is crucial to note that
 we would define the Boltzmann weights (BW) 
 of our vertex models 
 not by  the  $R$-matrix as conventional, but
 through the elements of the
     Lax operator (\re {nlslq2})
as  $L_{ab}^{j,k}(\la) =\omega_{a,j;b,k}(\la)$  by 
using 
matrix representations of the unifying q-deformed algebra (\re{nlslq2a}).

 We may
find an important
 representation of this algebra  through canonical fields as 
\be
 S^3=u, \ \ \   S^+=  e^{-i p}g_s(u),\ \ \ 
 S^-=  g_s(u)e^{i p}.
\ll{ilsg}\ee
with operator function
\be g_s (u)= \left ( {\kappa  }+\sin \al (s-u) (M^+ \sin \al (u+s+1)
+{M^- } \cos \al (u+s+1 ) ) \right )^{\ha}  { 1 \ov \sin \al } \ll{g}\ee
containing extra free parameters $\kappa$ and $s$. The unified model
represented by (\ref{nlslq2}) is a quantum integrable model and may be
considered as a
  {\it generalized lattice  SG model} through  realization  (\re{ilsg}),(
\re{g}). We are now in a position to generate the whole range of
integrable models, known as
well as new,  through various choices of the central elements
 $M^\pm$  as well as by mapping into different realizations.
In fact  (\re{ilsg}) can be 
directly mapped through the spin-$\ha$ operators $\vec \sigma$ or 
 realized  further 
in bosonic operators $\psi, \psi^\dag$
as 
$
\psi=e^{-i p} ({
 (s-u)})^{\ha},N=s-u,
$   etc. We may 
   repeat similarly the whole construction
at limit  $\al \to 0$  for
 generating the rational cases, or consider $q$ as roots of unity to get the
restricted models.
On the other hand at $\Delta \to 0$
we recover the   field models, while  
  $ \hbar  \to 0$   yields as usual 
 the  corresponding classical dynamical systems.  
As we show below, the choice for the inhomogeneity parameters
  $c_a^\pm$ can be of 
two  types: either as  constant parameters reproducing generally known
 models or as
  site $j$ dependent functions, which generates
   new classes of inhomogeneous or hybrid integrable  models.     

Motivated by the form of
(\re{ilsg})
  we find also a matrix representation of  
the unifying algebra (\re{nlslq2a}) 
as 
\be
<s,\bar m|S^3|m,s>=m \delta_{m,\bar m},  \quad <s,\bar m|S^\pm|m,s>
= f^\pm_s(m)\delta_{m\pm 1,\bar m} ,\ll{qsrep}\ee 
 with $f^+_s(m)=f^-_s(m+1) \equiv g_s(m) $
and using it construct
  the BW for our
  unified  vertex model
as
\be \omega_{\pm,k;\pm,k} (\la)=
{c_\pm^+} e^{i \al (\la \pm m)}+ {c_\pm^-} e^{-i \al (\la \pm m)}, \  
 \ \ \ 
\omega _{+,k;- ,k-1}=\omega _{-,k-1;+ ,k}=  
2 g_s(k-1) \sin \al, \ m=s+1-k
 , \ll{Avertex}\ee
where $k \in [1,D] $,
depends on the dimension $D$ of 
the matrix-representation of the  q-algebras.
Possible reductions of this unified model
(\re {Avertex}) would
 yield   new series of  vertex models (see FIG.1) with
 the familiar ice-rule is generalized here as the { color conservation} 
 $a+j=b+k$ for determining the nonzero  BW.
 
The eigenvalue solution of the transfer matrix related to the unified model
can  be found exactly 
through the algebraic Bethe ansatz, which therefore
would give also  exact solutions for
 all other  quantum and statistical models we construct here 
  in a unifying way.

\smallskip

\section {Construction of integrable   models }

\subsection{Trigonometric class with generic q}
 
\quad 
1.  The simplest constant choice  $c_a^\pm=\mp i$  giving 
$ M^-=0, \  M^+=2, $ reduces
 (\ref{nlslq2a})  
  to the well known
 $U_q(su(2))$   q-spin algebra  \c{jimbo}
$ [S^3,S^{\pm}] = \pm S^{\pm} , [ S^ {+}, S^{-} ] =  
[2 S^3]_q \equiv {\sin (2 \al S^3 \over \sin \al}.
$
and  (\ref{nlslq2}) to the corresponding Lax operator 
\be
L_{qspin}{(\la)} = \left( \begin{array}{c}
  [\la + S^3]_q \ \qquad \ \ 
  S^-   \\
    \quad  
   S^+  \  \qquad \ \  [\la - S^3]_q 
          \end{array}   \right), \quad \ll{qspin}\ee
expressed through the q-spin.
The simplest representation ${\vec S} = 
\ha  {\vec \si}$ reduces  (\ref{qspin}) further
 to (\re{R-mat}) and  recovers 
   the well known $XXZ$ {\it spin-$\ha$ chain} and also the 
6-vertex model as the related statistical system.

 On the other hand, for canonical
representation   (\ref{ilsg}) the form   (\re{g}) reduces to \\
$ g (u)={1 \ov 2 \sin \al}
  \left [ 1+ \cos \al (2 u+1)
 \right ]^\ha $
recovering the known integrable  
   {\it lattice sine-Gordon} model. However  for 
the corresponding statistical model
 the BW reduced from  (\re{Avertex})
takes the form
$ \omega _{\pm,j;\pm,j}(u)=
[u\pm m]_q, 
\ \ \omega _{+,j;- ,j-1}=\omega _{-,j-1;+ ,j}=  
 f^+_s(m) $ with 
$f^\pm_s(m)=([s\mp m]_q [s \pm m +1]_q ) ^{\ha}$, yielding 
a new series of integrable models, namely 
{\it q-spin 
$(4s+2)$-vertex model}, for different
 values of s (FIG.1b)). 

2. Another constant but different set of choices:
 $ \ c^+_1=c^-_2=1, \ \ c^+_2= c^-_1=0$  
   giving
  $ M^\pm=\pm \sqrt {\pm 1}$ 
   yield  an { exponentially} deformed  Lie
 algebra $
[ S^+, S^-]= {e^{2i\al S^3} \over 2 i \sin \al }
$
  The  representation (\re{ilsg})  
 is valid now with 
$  g(u)= {(1+e^{i \al(2 u+1)})^\ha \ov \sqrt {2} \sin \al} $
 and  reproduces
 the  {\it  lattice version of the 
quantum Liouville } model. One can as well construct the related vertex
model with the BW obtained as particular reductions of
 (\re{Avertex}) similar to the above case.

3. On the other hand another constant  
choice 
$ \ \ c^+_1=c^+_2=1, \ c^-_1= -{iq }  , \ c^-_2=  {i \over  q} 
$
leading to  $ M^+=2 {\sin \al } 
, \ M^-= 2 {\cos \al }  $
reproduces the well known 
$q$-boson algebra \c{qbos}
$
 [A,N] = A, \ \  [A^\dag,N] =- A^ \dag,\ \ \  [ A, A^ \dag ] = {\cos (\al (2N+1))
\ov \cos \al} 
$
from  (\ref{nlslq2a}) by denoting 
 $S^+= \rho A, \ S^-= \rho A^\dag, \ S^3= -N, \ \
 \rho=  ( \cot \al)^\ha
$
and corresponds to a  new quantum integrable
 {\it $q$-bosonic  model} with the Lax operator reduced from (\ref{nlslq2})
as
\be
L_{qboson}{(\la)} = \left( \begin{array}{c}
  e^{i \al \phi} [\la -(\hat N+\phi)]_q \ \qquad \ \ 
  \kappa A^\dag  \\
    \quad  
  \kappa A \  \qquad \ \  e^{-i \al \phi} [\la +(\hat N+\phi)]_q 
          \end{array}   \right), \quad \ll{qbosonL}\ee 

Realizing the $q$-boson  
 through standard boson 
 one may construct   
   an   integrable quantum 
 model,   representing 
 a lattice version  
 of the quantum {\it derivative nonlinear Schr\"odinger equation} (QDNLS)
   and  at the continuum limit the corresponding field
model.
 Fusing two such models one can
build further  a quantum integrable
 {\it massive Thirring}
 model \c{qdnls}.
The QDNLS
 is also  related to the exactly solvable interacting bose gas model with
derivative $\delta$-function potential.

For constructing the related vertex model we require 
  matrix
representation of the q-bosonic operators,
 which   with the present choice of the central elements
and   assuming  $\kappa =s=0, \ n=-m$, may be derived
  from 
(\re{qsrep}) as
$<\bar n|N|n>=n \delta_{n,\bar n}, \ \  \quad < \bar n|A^\dag |n>
= f_0^-(n)\delta_{n+1,\bar n} ,
\quad < \bar n|A|n>
= f_0^+(n)\delta_{n-1,\bar n}$ with 
$f_0^-(n)=([1+n]_q[[-n-1]]_q)^\ha= {1 \ov \sqrt 2}[1+n]^\ha_{q^2}, \ 
f_0^+(n)=f^-(n-1)={1 \ov \sqrt 2}[n]_{q^2}^\ha $. 
  Consequently the BW of the related statistical model 
 reduced from  (\re{Avertex}) take the form   
as
 \be \omega_{\pm,j;\pm,j}(u)=
 ie^{\pm i \al  \phi} 
[u \mp (j+\phi -1)]_q ,\ \ 
\omega _{+,j;- ,j-1}=\omega _{-,j-1;+ ,j}=
 f^+(j-1)= {1 \ov \sqrt 2} [j-1]^\ha_{q^2} ,\ll{qbboltzman}\ee
with $\phi = \ha (1+ {\pi \ov 2 \al}) $
which generates new {\it q-boson $(4n+2)$-vertex model} (FIG.1c)).

4. It is easy to see that for all the following  parameter choices:
$ i) \ \ c_a^+=1 \ , a=1,2, \ \mbox{or} \ \
 ii)\ \  c_1^\mp=\pm 1  , \  \ \mbox{or} \ \  iii)\ \  c_1^+=1, \
$
with   rest of the  $c'$s being zero,   
we get $M^\pm=0$
 with the   underlying  algebra reducing to
$[ S^+, S^-]= 0, \ [ S^3, S^\pm]= \pm S^\pm $
and (\ref{ilsg})  simply to  
$ S^3 =-ip, \ S^\pm=  \al e^{\mp u } \ \ 
.$ 
This yields the  {\it  relativistic
quantum Toda chain} recovering also its different Lax operator constructions.

The statistical systems related to this case however
 seem to be give uninteresting
vertex models with infinite
matrix traces.

\subsection{Models with  q roots of unity}

It is obvious that the integrability of a system remains intact even if
  $q$ is chosen as  solutions of  $q^p=1$.
with parameter $\al$ taking discrete values $\al _a =2 \pi {a \ov p}, a= 1,2,
\ldots,p-1$ \c{saclay}.However  this  opens up an excellent possibility for
regulating the dimension of the  representation of the underlying algebra.
All quantum models constructed above will yield 
 their restricted versions
when q is taken as the roots of unity. Among these models 
only a few appear to have been studied earlier, which include 
 {\it restricted sine-Gordon
model}  connected with the q-spin and  {\it restricted DNLS}
 associated with the q-boson.

The related vertex models  
 generate now completely new series, 
since the number of the possible configurations can be regulated  
through  integer p.
To analyze this fact we focus on the action of $S^-$ in  (\re{qsrep})
assuming
$\kappa=0$ : 
and  observe  that
   due to $[p]_q=\sin \al_a p= 0,$  unlike generic q we can get now  
$S^-|-\bar s,s> = 0$ 
 at $ \bar s= p-(s+1)$,  which
  reduces matrix (\re{qsrep}) to a finite dimensional representation. As a
consequence, for q-spin with fixed $p, 0<p<2s+1,$ we get now $p-1$ number of
different
 $ (4p-2)$-vertex models for different discrete values of $\al _a$.
 
The situation  becomes  more interesting when applied to the  
q-boson   with finite $p$,
 since  now in place of its unbounded representation 
we obtain a finite dimensional  matrix representation 
 leading to 
 an intriguing series of $(2p-2)$-vertex models with
 BW described by the same form (\re{qbboltzman}) as for the generic
q-boson case, but now  with
different possible  discrete parameter values
$q=e^{i \al_a }, a=1, 2, \ldots, p-1$.
\subsection{Rational class}
At the limit  $\al \to 0$ or $q \rw 1$  
 the  Lax operator   (\ref{nlslq2}) , the BM 
 (\re{Avertex})  as well as the  $R$-matrix reduce to their
corresponding rational limits with the underlying algebra 
becoming
\be  [ s^+ , s^- ]
=  2m^+ s^3 +m^-,\ \ \ \ 
  ~ [s^3, s^\pm]  = \pm s^\pm , \ \ [m^\pm ,\cdot]=0,  
  \ll{k-alg} \ee
with the central elements $M^\pm \to m^\pm$.
We can  find as before  
various realization of this algebra
 directly from (\re{ilsg}) 
  using however a limiting form of (\re{g}) as
$g_0(u)=i ( (s-u)(m^+ (u+s+1)+m^-))^\ha$ and  construct 
 in the similar way as above different integrable quantum and
statistical models belonging to this rational class. 

1.Thus  at  
 $  m^+  =  1,m^-  =  0 $, when  one gets the standard 
   $su(2)$ algebra,
the  quantum models produced are  the $XXX$ {\it  spin chain}
and
the  {\it lattice   NLS} model, while  the corresponding integrable vertex
models 
related to the spin-s operators  recover those obtained earlier through
fusion technique \c{babus}.

2.A complementary  choice
 $  m^+  =  0,m^-  =  1, $ on the other hand 
 yields the bosonic algebra 
generates  another quantum  {\it simple   
lattice NLS} model
\c{kunrag}. The corresponding  bosonic
vertex model  represents a nontrivial integrable statistical
 model, apparently never studied before.

3.A trivial choice  $m^\pm  = 0$  reproduces
  the  algebra related to the relativistic Toda chain considered above, 
which however in  the rational  limit gives    
     its well known   {\it nonrelativistic} variant.
 The corresponding vertex model as before seem to be not
physically interesting.

\section{Inhomogeneous and hybrid models}

 An immediate generalization of all the above models is possible 
   by considering 
the central elements  $c'$s appearing in the Lax operator (\ref{nlslq2})
to be  site dependent functions.
This would lead to a  new class of integrable   
 inhomogeneous
 extensions of the above models and may be interpreted as models with
impurities, varying external fields, incommensuration etc.

Thus we can construct for example 
  variable mass sine-Gordon, Liouville models, variable coefficient NLS ,
Toda chain models etc similarly the corresponding integrable q-spin and
q-boson vertex models in varying external field.


Another intriguing  class of models can be formed
by regulating 
  the inhomogeneity such that the   
 Lax operators of   different models are arranged to sit 
 at different lattice sites along the chain defining 
the  transfer matrix of the model as
$   \tau (u)= tr\left(\prod_\beta\prod_{j=1}^{N^{ (\beta)}}
L_j^{(\beta)}(u)\right), $ where $L_j^{(\beta)}(u) $ indicates Lax operators
of different  models belonging to the same class with
 the same $R$-matrix and occurring 
$N^{(\beta)} $ times in the total number of sites $N=\sum_\beta N^{(\beta)} $.

Thus one can construct new series of exotic integrable  models like 
 hybrid 
 sine-Gordon-Liouville model,  hybrid NLS-Toda chain  or
 spin-boson model,  describing 
 different types of nonlinear interactions 
at different domains of the coordinate space. Similarly one obtains 
hybrid integrable statistical 
 models like (un-)deformed spin-boson models etc. 
combining different types of vertex models belonging to the same class
(FIG.1).

Considering the continuum limit
$\Delta \rw 0$  we recover  the quantum field models
    from their respective  lattice versions constructed above,
 while their field Lax operator ${\cal L}(x, \la) $ takes the form 
$L_j(\la) \rw I+ i\De {\cal L}(x, \la) +O(\De^2).
$ The
associated $R$-matrix  however remains the same as its discrete
counterpart.
 At the limit $\hbar \rw 0$ one obtains the corresponding 
classical models 
with the $R$-matrix  reducing also  to its classical form 
 $R(\la)=I +\hbar r(\la)
+O(\hbar^2)$. For statistical models on the other hand
the most   relevant is the thermodynamic limit: $N \rw \infty$,  with  
$\Delta, \hbar$ fixed.

\section { Unified  solutions}:
There is a well formulated algebraic Bethe ansatz method
for exactly solving the 
  eigenvalue problem of the transfer matrix 
$\tau(u)=tr (\prod_l^N L_l(u))$, when the $L,R$ matrices are given \c{qist}.
Therefore for our   unified   model represented by 
  the generalized Lax operator 
 (\ref{nlslq2}) and the $R$-matrix (\re{R-mat}) we get
\be
{\Lambda}(\la)=  
(<0|\hat L^{11}(u)|0>)^N(u)\prod_k^n {f}(u_k-u)+
(<0|\hat L^{22}(u)|0>)^N(u)\prod_k^n {f}(u-u_k)
, \ \ f(u)= { [u+ 1]_q \ov [u]_q}. 
\ll{ev}\ee
with all possible solutions of $\{u_k\}$ to be  determined from 
 the Bethe equations \be
\left ({(<0|\hat L^{11}(u_j)|0>)\over (<0|\hat L^{22}(u_j)|0>)}\right )^N=
\prod_{k \neq j}^n {[u_j-u_k+1]_q \over [u_j-u_k-1]_q }, \ \  \ j=
1,2,\ldots, n.
\ll{be}\ee 
Here $|0>$ is the pseudo-vacuum and 
the only model dependent parts in both the above equations  are given by 
   the actions of the upper and lower
diagonal operator elements $L^{11},L^{22} $ of the Lax operator.
Note that since in our scheme the $R$-matrix is the same for all models 
and their $L$-operators are given through various reductions of 
 (\ref{nlslq2}), 
the eigenvalue form (\re{ev}) together with the Bethe equation (\re{be}) 
give  
 a unifying   scheme for  exactly solving all the  integrable
quantum as well as  statistical models 
constructed here.

For our vertex models  the Lax operator elements 
 in the above equations  
should be replaced by their matrix representations expressed through 
the  BW ({\re{Avertex}) as
$ 
(<0|\hat L^{11}(u)|0>)= \omega_{+,1;+,1}
(<0|\hat L^{22}(u)|0>)=\omega_{-,1;-,1}
$.
 The
total number of independent solutions for the eigenstates should be equal to
the dimension of the vector space on which the transfer matrix acts.
For the vertex models in our construction this is
 $K=D^N $ and   the partition  function 
 should be given through  exact eigenvalue solutions  as 
  $Z=lim_{M,N \to \infty}tr_v(\tau^M)=
lim_{M,N \to \infty}\sum_k^K \Lambda_k^M$, in the thermodynamic limit.
 Note that 
though in general the  dimension of the transfer matrix may even 
  be infinite (including degeneracies),
the corresponding   partition function must be  well defined.
The  Hamiltonian of the  quantum models related to such statistical
systems would    generally  involve 
nonlocal 
 interactions which however are not relevant for the  associated
  vertex models we
are concerned with.
Though the Bethe equation gives the form for deriving exact solutions, only in the
thermodynamic limit one  usually  expects to  find such solutions
 by converting
 this algebraic equation 
 into an integral equation \c{6v}. For   our unified 
model assuming $c^\pm_2={r_2 \ov r_1}{(c^\pm_1)}^*$,
  for physical reasons we can 
similarly
derive 
from (\re{be}) 
 \be
V(c^\pm_1,u)=2 \pi \rho(u)-\int dv \rho (v) {\sin 2\al s \ov 
\cos 2\al s -cosh (u-v)}.
\ll{bbe}\ee 
It is important to note that the rhs of the above
equation totally coincides  with that for the 6-vertex model \c{6v}, 
while the lhs  represents the model-dependent part and
is expressed through  BM $ \omega_{+,1;+,1}=
c^+_1e^{u+is \al}+c^-_1e^{-(u+is \al)}\equiv  r_1  \omega(u) $
given by (\re{Avertex})  as
$V(c^\pm_1,u)= {\omega^{'}\omega^{*}-\omega \omega^{*'}\ov | \omega|^2}$.
Therefore  the integral
 equations related to  all vertex models including 
the q-spin and q-boson vertex models  can be obtained 
 from the single equation  (\re{bbe}) at proper choices 
 of $c^\pm_1 $, which should also  naturally include 
  the well known case of  
$6$-vertex model \c{6v}.
Hybrid models, models with q roots of unity or at $q \to 1$
can also be covered  by the above unifying scheme or its extensions.

\section {Concluding remarks}

Thus we have    prescribed  an unifying scheme for 
 constructing as well as solving integrable  quantum
and vertex models of certain classes 
, which covers  lattice and field models
of  (non-)relativistic and (isotropic)anisotropic types as well as the
  corresponding
 integrable statistical models.
Along with the known models   one can construct  new
models, including a novel series
of  q-spin and q-boson vertex models related to q at roots of
unity. 
 Inhomogeneous and hybrid  models  constitute new exotic classes of
integrable models.
 Using algebraic Bethe ansatz we can also
 find exact solutions for  all these
models in a unifying way. For some other details concerning our unified
scheme the readers may consult the related works \c{1,2,3,4}.

\ni {\bf Acknowledgment}: 

I thank Profs. Bernard Nienhuis, Doochul Kim, Deepak dhar and Ashok
Chatterjee for stimulating discussions.
\\ \\
{\small  FIG. 1. Integrable vertex models with horizontal (h) links taking $2$ values,
while the vertical (v) ones may  have $D$ possible values.  
a) 6-vertex   b) q-spin vertex  and  c) q-boson vertex models. 
Combining a,b,c) an integrable hybrid model may be formed.  $q^p=1$ gives
 $D=p$ in b) and c)}

\newpage

 
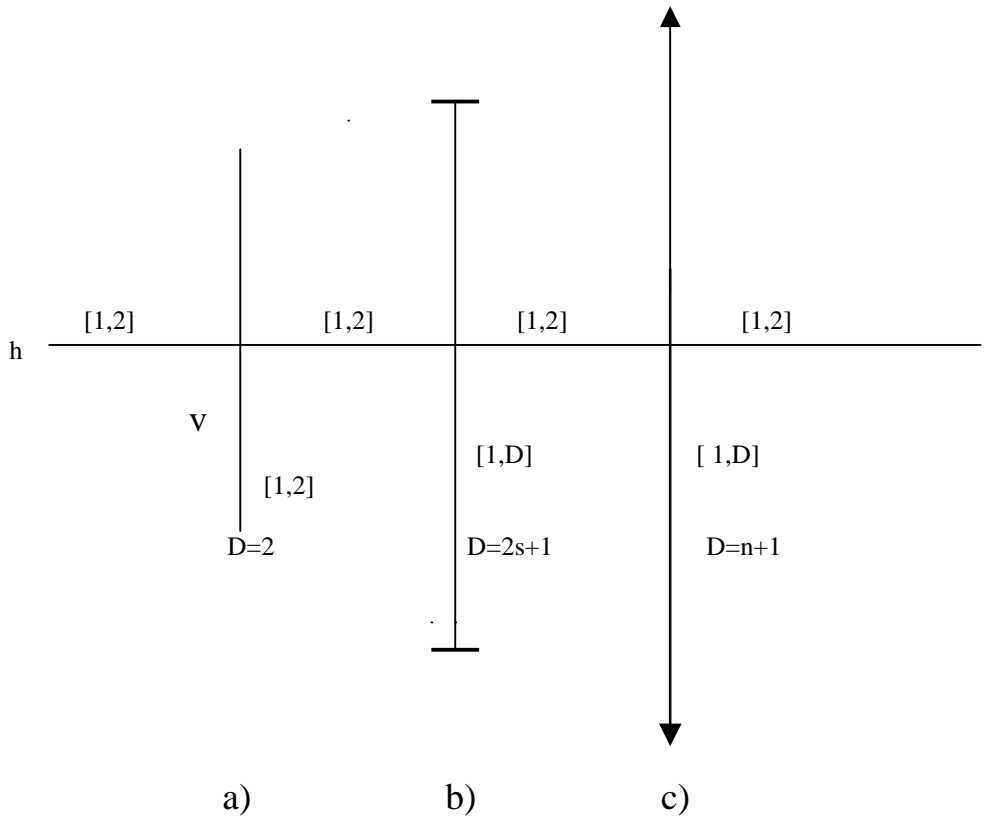
\end{document}